\documentclass[aps,prl,twocolumn,
showpacs,groupedaddress,floatfix]{revtex4}
\usepackage{graphicx,dcolumn,bm}
\usepackage{amsfonts,amsmath,amssymb,amsthm,amscd}
\begin{document}
\title{Collapse and revival of electromagnetic cascades in focused intense laser pulses}
\author{A.~A. Mironov}\email{mironov.hep@gmail.com}
\author{N.~B. Narozhny}
\author{A.~M. Fedotov}
\affiliation{National Research Nuclear University MEPhI (Moscow Engineering Physics Institute), 115409 Moscow, Russia}
\date{June 2, 2014}

\begin{abstract}
We consider interaction of a high-energy electron beam with two counterpropagating femtosecond laser pulses. Nonlinear Compton scattering and electron-positron pair production by the emitted photons result in development of an electromagnetic ``shower-type'' cascade, which however  collapses rather quickly due to energy losses by secondary particles. Nevertheless, the laser field accelerates the low-energy electrons and positrons trapped in the focal region, thus giving rise to development of electromagnetic cascade of another type (``avalanche-type''). This effect of cascade collapse and revival can be observed at the electron beam energy of the order of several GeV and intensity of the colliding laser pulses of the level of $10^{24}$W/cm$^2$. This means that it can be readily observed at the novel laser facilities which are either planned for the nearest future, or are already under construction. The proposed experimental setup provides the most realistic and promissory way to observe the ``avalanche-type'' cascades.
\end{abstract}

%52.27.Ep Electron-positron plasmas
%52.50.Dg Plasma sources
%41.75.Jv Laser-driven acceleration
%42.50.Ct Quantum description of interaction of light and matter; related experiments
\maketitle

Amazing progress in elaboration of high-power laser technologies was demonstrated  in recent decades. By now, laser intensity has already surpassed the level of $10^{22}\text{W/cm}^2$ \cite{Yanovsky}. Recently, several projects aiming at intensities up to $10^{26}\text{W/cm}^2$ were announced and are currently under active development, see, e.g., \cite{ELI,XCELS}. When realized, they would open unique possibilities to observe a variety of yet unexplored exotic QED effects of nonlinear interaction of electromagnetic radiation with matter and vacuum, see reviews \cite{Review1,Review2} for the details.

Experimental tests of Intense Field QED were pioneered almost two decades ago at SLAC \cite{SLAC}, where both the nonlinear Compton effect and the nonlinear Breit-Wheeler process have been observed for the first time. The forthcoming facilities \cite{ELI,XCELS} would allow to reproduce the SLAC-like experiments at a principally new level. Instead of rare events of photon emission and $e^+e^-$-pair creation, the long chains of sequential events of those processes (cascades) would be observed \cite{Bulanov}, see also \cite{Sokolov}.

The cascades discussed in Refs.~\cite{Bulanov,Sokolov} are very similar to the well studied air showers which are induced by cosmic rays in atmosphere \cite{Auger, Gaisser}. For the sake of brevity we will call them hereinafter \textit{S(Shower)-type cascades}. However, another mechanism of cascade development is much more intriguing. The point is that a cascade may occur due to acceleration of charged particles by the laser field and can be seeded even by a particle initially at rest \cite{Kirk1, Kirk2, Laser-limit, Elkina-STAB}. Such cascades have a lot of similarities with the Townsend discharge in gases \cite{Townsend}, or the avalanche breakdown in insulators or semiconductors, see, e.g., \cite{Av}. We will call them here \textit{A(Avalanche)-type cascades}. It is worth noting that since an A-type cascade withdraws the energy necessary for its development from the field, it may result in substantial depletion of the laser field \cite{Kostyukov} and, very likely, even bound the maximally attainable value of laser intensity \cite{Laser-limit}.

In the case of S-type cascades, creation of $e^+e^-$ pairs and hard photons occurs at the expense of kinetic energy of the initial and secondary particles. The multiplication factor for such a cascade depends also on the laser field strength. The higher is the field strength, the higher are the probabilities of particle creation processes and, hence, the multiplication factor. Nevertheless, the S-type cascade collapses at any laser intensity when the energy of secondary particles decreases to some threshold value. On the other hand, occurrence of the A-type cascades does not necessary requires high energy of a primary particle. However, the presence of a very strong electric field is absolutely necessary to ensure fast energy gain for a particle initially at rest \cite{Laser-limit}.

The goal of this paper is to show that it is possible to observe both types of cascades in the same SLAC-like experiment at the forthcoming facilities \cite{ELI,XCELS}. This could be possible if the time of collapse for an S-type cascade were small compared to duration time of the laser pulse $\tau_L$. Then, the mechanism of acceleration of relatively slow particles, which did not have enough time to leave the focal region, could turn on giving start of an A-type cascade development. In such a way we could encounter revival of the cascade process.

We will consider an electron beam colliding with the field of two counterpopagating circularly polarized intense focused laser pulses with frequency $\omega$ and duration $\tau_L\gg 1/\omega$, so that magnetic component of the field vanishes in the focus of the standing wave and the field there can be considered as a rotating electric field, compare \cite{Laser-limit, Elkina-STAB}. As in Ref.~\cite{Elkina-STAB}, we assume that for both pulses $E,H\ll E_S$, where $E_S=\frac{m^2c^3}{e\hbar}=1.32\times 10^{16}$V/cm is the QED critical field, and parameter $\xi=e\sqrt{-A^\mu A_\mu}/mc\gg 1$,  where $A_\mu$ is the field 4-potential. We assume that the direction of the electron beam coincides with the direction of one of the laser pulses and the initial energy of the electrons $\varepsilon_0\gg mc^2$.

For the case of a laser field of optical frequency and ultra-relativistic particles one can use probability rates for photon emission $W_{\gamma}$ and $e^+e^-$-pair creation $W_{cr}$ in the approximation of a locally constant crossed field \cite{Laser-limit}. Then, they are exclusively determined by the dynamical quantum parameters of participating particles
\begin{equation}
\label{eq:chi}
\chi_{e,\gamma} = \frac{e\hbar}{m^3c^4}\sqrt{\left(p_0\mathbf{E}+\mathbf{p}\times\mathbf{H}\right)^2-\left(\mathbf{pE}\right)^2},
\end{equation}
and can be estimated as
\begin{equation}
\label{eq:W}
W_{\gamma,cr}\sim\frac{\alpha m^2c^4}{\hbar \varepsilon_{e,\gamma}}\chi_{e,\gamma}^{2/3},\;\chi_{e,\gamma}\gtrsim 1,
\end{equation}
\cite{probabilities}, where 4-momenta $p^\mu$ are: $p^\mu_e=(\varepsilon_e,\mathbf{p}_e)$ for electron or positron, and $k^\mu_\gamma=(\varepsilon_\gamma=\hbar\omega,\mathbf{k}_\gamma)$ for a photon, $\mathbf{E}$ and $\mathbf{H}$ are the local values of the electric and magnetic field. For small values of parameters (\ref{eq:chi}), $W_\gamma\sim\frac{\alpha m^2c^4}{\hbar \varepsilon_{e}}\chi_{e}$, while the probability of pair creation is suppressed exponentially $W_{cr}\sim\frac{\alpha m^2c^4}{\hbar \varepsilon_{\gamma}}\chi_{\gamma}\exp(-8/3\chi_\gamma) $.

Let us estimate duration of the S-type cascade $\tau_S$ first. The initial value of parameter (\ref{eq:chi}) for a primary ultrarelativistic electron can be estimated as $\chi_0\sim \frac{\varepsilon_0}{mc^2}\frac{E_0}{E_S}$, where $E_0$ is the peak value of the electric field strength in the focal region of the cumulative field of two colliding pulses. Any involved particle transforms into two in every event, so that multiplicity of the S-type cascade can be estimated as in Ref.~\cite{Akheizer},
\begin{equation}
\label{eq:2n}
2^n=\frac{\chi_0}{\chi_f},
\end{equation}
where $n$ is the number of generations of secondary particles and $\chi_f$ is the value of parameter (\ref{eq:chi}) for the final electrons. Thus, for $\tau_S$ we have $\tau_S\sim t_en$, where $t_e\sim W^{-1}_\gamma(\varepsilon_0, \chi_0)$ is the mean lifetime of a primary electron with respect to hard photon emission. Finally,
\begin{equation}
\label{col_time}
\tau_S\sim \tau_C\frac{\varepsilon_0}{\alpha mc^2}\chi_0^{-2/3}\log_2\left(\frac{\chi_0}{\chi_f}\right),
\end{equation}
where $\tau_C=\hbar/mc^2$ is the Compton time. The primary electrons are supposed to be ultrarelativistic, so that $\chi_0>1$. Obviously, the S-type cascade collapses when $\chi_f<1$. We will choose $\chi_f$ to be $\sim 0.1$. At such $\chi_f$, the development of the S-cascade cannot continue since the mean lifetime of a photon with respect to pair production, $t_\gamma\sim W^{-1}_{cr}\sim \exp(8/3\chi_f)$, becomes exponentially large.

We are interested in the case when $n>1$ and the collapse time of the S-type cascade is less than the laser pulse duration $\tau_L$. Hence, the following requirements
\begin{equation}
\label{req}
t_{e,\gamma}<\tau_C\frac{\varepsilon_0}{\alpha mc^2}\chi_0^{-2/3}\log_2(10\chi_0)<\tau_L,
\end{equation}
should be respected. Let $\varepsilon_0=3$ GeV and $E_0=3.2\times10^{-3}E_S$. Then $\chi_0\approx20$, and the left condition in (\ref{req}) is satisfied at the expense of large logarithm $\log_2(10\chi_0)\approx8$. The right condition in (\ref{req}) is satisfied for $\tau_L\gtrsim10$ fs.

It was shown in Refs.~\cite{Laser-limit,Elkina-STAB} that a charged particle acquires very large transverse, with respect to the direction of the laser pulses propagation, momentum under the action of rotating electric field, and the dynamical parameter $\chi$ of the particle gains the increment $\Delta\chi\sim1$ within a small fraction $t_{acc}\sim\frac{E_S}{E_0}\sqrt{\frac{mc^2}{\hbar\omega}}\tau_C$  of the rotation period, $\omega t_{acc}\ll1$. If trajectory of the particle contorts so strongly that it can emit a hard photon in the direction transverse with respect to its initial propagation, which by-turn can create a pair, the A-type cascade occurs. In our case, the initial particle has a large longitudinal momentum and accordingly large value of parameter $\chi$. Therefore a noticeable contortion of the particle trajectory necessary for development of the A-type cascade will occur only if $\chi\sim1$.

Let us explain this issue in more details. The momentum of a particle moving in the field of two colliding laser pulses  can be represented as $\mathbf{p}(t)=\mathbf{p}_{\parallel}(t)+\mathbf{p}_{\perp}(t)$. The longitudinal component of the momentum $\mathbf{p}_{\parallel}(t)$ varies only due to emission of photons, and its characteristic variation time is $t_{\parallel}\sim\tau_S\lesssim\tau_L$. Variation of the transverse momentum component $\mathbf{p}_{\perp}(t)$ is determined by the rotating electric field $\mathbf{E}(t)=\{E_0\cos(\omega t), E_0\sin(\omega t),0\}$,
$$\dot{\mathbf{p}}_{\perp}(t)=e\mathbf{E}(t)\,,$$
and thus its characteristic variation time is $t_{\perp}\sim 1/\omega$. We assume that a charged particle arrives at the focal region at the moment $t=0$ with zero transverse momentum, $\mathbf{p}_{\perp}(0)=0$. Then taking into account the relation $t_\parallel/t_\perp\sim\omega\tau_L\gg1$, we can calculate $\chi(t)$ according to Eq.~(\ref{eq:chi}) and  get the formula
\begin{equation}\label{chi*}
  \chi(t)\approx \sqrt{\chi^2_\parallel(t)+(\Delta\chi_\perp(t))^2},
  \end{equation}
valid within the time interval $\Delta t\lesssim 1/\omega$. Here $\chi_\parallel=\chi(0)=\frac{E_0}{E_S}\sqrt{1+\frac{p^2_{\parallel}}{m^2}}$ has the meaning of the dynamical parameter in the absence of accelerating field, and $\Delta\chi_\perp=2\xi\frac{E_0}{E_S}\sin^2(\frac{\omega t}{2})$, compare to Eq.~(9b) in Ref.~\cite{Elkina-STAB}. $\Delta\chi_\perp$ gains the value $\sim1$ within $t_{acc}\ll 1/\omega$. This can essentially contort the trajectory and thus give start to an A-type cascade only if $\chi_\parallel\sim1$, as it is seen from Eq.~(\ref{chi*}).

So, the A-type cascade will retard with respect to the S-type cascade start and the delay time $\tau_R$ is determined by Eq.~(\ref{col_time}) with $\chi_f=1$. Consequently, we can estimate the duration of the A-type cascade $\tau_A$ as $\tau_A\sim\tau_L -\tau_R$. Certainly, the conditions $t_{acc}\ll t_{e,\gamma}<\tau_A$ must be respected. It is easy to check that they hold for the values of $\varepsilon_0, E_0$ and $\tau_L$ adopted above.

For more accurate analysis we use Monte-Carlo simulations. We assume $e^-$ and $e^+$ are moving classically between the acts of emission and solve relativistic equations of motion numerically. Timepoints of the quantum events are determined in a manner similar to Refs.~\cite{GEANT4, Duclous-Kirk}. The code was tested by simulating cascades initiated by high-energy electrons in a constant homogeneous transverse magnetic field and by simulating cascades caused by electron seeded at rest in the uniformly rotating homogeneous electric field. The results are in reasonable agreement with the cascade profiles from Refs.~\cite{Anguelov} and \cite{Elkina-STAB} respectively.

In our simulations we use a realistic model of circularly $e$-polarized focused laser pulses \cite{Narozhny-Fofanov} with gaussian temporal amplitude envelopes $E\propto \exp[-4(t\mp z)^2/\tau_L^2]$. Frequency of the laser pulse is $\hbar\omega=1$eV, duration $\tau_L=10$fs, and focusing parameter $\Delta=0.1$. Laser pulses propagate along and against $z-$axis and are focused at $z=0$ at the moment $t=0$. Initial parameters of electron beam are chosen so that without laser field electrons move along $z$-axis and reach $z=0$ at $t=0$. The initial energy of electrons $\varepsilon_0=3$ GeV while the peak value of electric field strength $E_0$ was varied in numerical experiments. The results were averaged over $10^3$ simulation runs for each choice of parameters.

\begin{figure}
	\includegraphics[width=0.99\columnwidth]{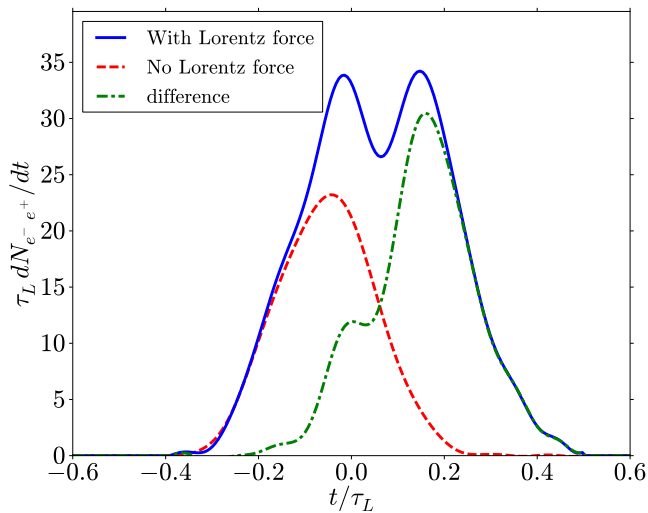}
	\caption{(Color online) Pair creation rate versus time. Solid line is the result of simulation, dashed line --- simulation in the absence of Lorentz force, dot-dashed line --- difference between the solid and the dashed curves. Initial parameters: $E_0=3.2\times 10^{-3}E_S$, $\varepsilon_0=3$ GeV.}
	\label{fig:dN}
\end{figure}
\begin{figure}
	\includegraphics[width=0.99\columnwidth]{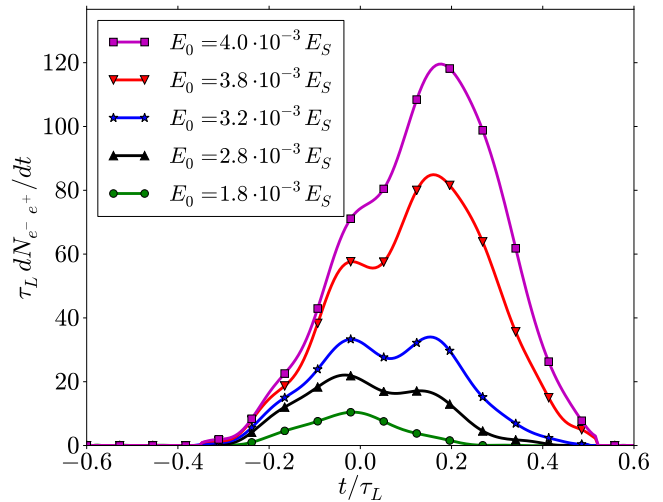}
	\caption{(Color online) Pair creation rate versus time for different values of $E_0$. The initial electron energy is $\varepsilon_0=3$ GeV.}
	\label{fig:dN_E}
\end{figure}

\begin{figure}
	\includegraphics[width=0.99\columnwidth]{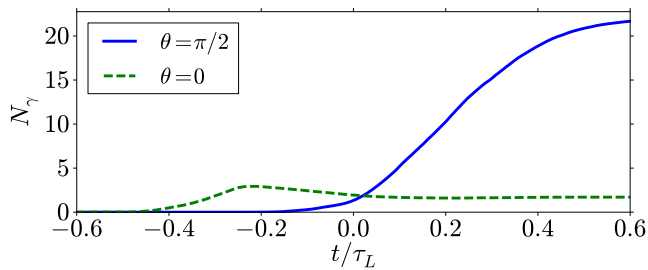}
	\caption{(Color online) The total number of photons emitted in selected directions versus time. Dashed line -- the number of photons emitted along the direction of the initial electron beam ($\theta\in[0,\,0.1]$ rad); solid line -- the same for transverse direction ($\theta\in[\frac{\pi}{2}-0.05,\,\frac{\pi}{2}+0.05]$ rad). $E_0=3.2\times 10^{-3}E_S$, $\varepsilon_0=3$ GeV.}
	\label{fig:theta_dir}
\end{figure}

	
%\begin{figure}
%	\includegraphics[width=0.99\columnwidth]{dN_gdt.eps}
%	\caption{(Color online) Photon production rate versus time. Solid line corresponds to result of simulation, dashed line --- simulation without Lorenz force, dot-dashed line --- difference of solid and dashed curves. Initial parameters: $E=1.6\times 10^{-3}E_c$, $\varepsilon_0=3$ GeV.}
%	\label{fig:dNg}
%\end{figure}
The results of the simulations are presented in Figs. \ref{fig:dN}-\ref{fig:theta_dir}. All distributions are normalized assuming that cascading was initiated by a single $e^-$. Pair creation rate $dN_{e^-e^+}(t)/dt$ versus time is shown in Fig. \ref{fig:dN} for $E_0=3.2\times 10^{-3}E_S$, $\varepsilon_0=3$GeV. Suppose the Lorentz force is ``turned off'' and electrons are not accelerated by the field. Then they lose energy only due to photon emission and only S-type cascade takes place. Pair creation rate for the case with the ``turned off'' Lorentz force is represented by the dashed line. The cascading starts immediately after electrons meet the counterpropagating laser pulse and collapses in the time interval $\tau_S\approx 0.6\tau_L$. The solid line shows the total rate. We see that at the initial stage it coincides with the dashed line and this means that in the beginning we have only the S-type cascade. In full agreement with our estimates, after approximately $0.3\tau_L$ the total rate begins to exceed the ``no-Lorentz force'' rate. Finally, we see the second peak of the solid line signifying revival of the cascading process due to the course of the A-type cascade. The dot-dashed line on Fig. \ref{fig:dN} shows the difference between the total and ``no-Lorentz force'' rates. It corresponds to the A-type cascade which develops due to acceleration of electrons in transverse direction by the laser field.

Pair creation rates versus time are presented in Fig.~\ref{fig:dN_E} for different values of the field strength $E_0$. One can see that the
revival of cascading (represented by the second peak) occurs only for the strong enough field. For the chosen values of laser and electron beam parameters it approximately equals to $E_0=2.8\times 10^{-3}E_S$.

Another signature of cascade revival may be found in angular distributions of photons. Ultrarelativistic electrons initiating the S-type cascade emit photons along the direction of their propagation. The A-type cascade arises only after occurrence of fast charged particles accelerated in transverse direction. These particles will predominantly emit photons also in transverse direction. In Fig. \ref{fig:theta_dir} one can see the total number of photons $N_\gamma$, emitted at different angles $\theta$ with respect to the direction of propagation of the initial electron beam, versus time. The number of photons emitted at $\theta=0$ first increases at $t<0$ and then begins to decrease because the S-type cascade collapses but some of the earlier emitted photons can still create pairs at $t>0$. At the same time, the number of photons emitted at $\theta=\pi/2$ starts to increase significantly at $t\geq0$. The times of collapse of the S-type and onset of the A-type cascades are in good agreement with the results presented in Fig.\ref{fig:dN}. The substantial difference in the numbers of photons emitted in these two directions can be explained by the numbers of particles involved. The S-type cascade is initiated by a single electron, while much greater number of generated secondary particles participate in the development of the A-type cascade.

%Multiplicity of the S-type cascades is limited by the initial energy of the electron beam and by the laser field strength. The number of pairs created in the course of the S-type cascade can be estimated according with Eq.~(\ref{eq:2n}) as $n\sim\log_2(10\chi_0)$. If the A-cascade occur, the total number of pairs $N_{e^-e^+}$ will exceed this number significantly. This is illustrated by Fig. \ref{fig:N_E}. At low laser field strength, when the A-cascade does not arise, $N_{e^-e^+}$ practically coincides with the estimate (\ref{eq:2n}), but at higher $E_0$ it exceeds the estimate significantly. This can be treated as a signature of the presence of the A-type cascade.

%\begin{figure}
%	\includegraphics[width=0.99\columnwidth]{N(E).eps}
%	\caption{(Color online) Total number of pairs versus peak value of laser field strength. Solid line corresponds to simulation. Green dashed line -%-- estimated number of pairs by Eq. (\ref{eq:2n}). Initial energy of electron beam $\varepsilon_0=3$ GeV.}
	%\label{fig:N_E}
%\end{figure}

To conclude, we predict the effect of collapse and revival of QED cascading. This effect can be observed in same experiment when a beam of ultrarelativistic electrons collides with an intense laser field. In the setup considered in this letter, the laser field was formed by two counterpropagating circularly polarized laser pulses of optical frequency. We have demonstrated that cascade of Shower-type arises first, and then it collapses rather quickly due to energy losses by the secondary particles. However, when the energy of the secondary charged particles becomes small enough, the mechanism of their acceleration by the laser field turns on. This initiates the development of the avalanche-type cascade and leads to revival of cascading \textit{in toto}. We have shown that the maxima of pair creation rates for two mechanisms of cascading can be distinguished for the chosen setup at the peak value of the field strength $E_0\approx 2.8\times10^{-3}E_S$.  This roughly corresponds to the laser intensity of each pulse $I\approx10^{24}\,\mathrm{W/cm}^2$, which is expected to be obtained in the nearest future \cite{ELI, XCELS}. The proposed experimental setup provides the most realistic and promissory way to observe the avalanche type cascade.

\begin{acknowledgements}
The authors are grateful to G. Mourou, A. Di Piazza, A. Ilderton, H. Takabe and S.P. Kim for helpful discussion and valuable remarks.
The research was supported by the Russian Fund for Basic Research (Grant No. 13-02-00372) and the President Grant for Government Support of Young Russian Scientists and the Leading Scientific Schools of the Russian Federation (Grant No. NSh-4829.2014.2).
\end{acknowledgements}

\end{document}